\documentclass[aps,onecolumn,superscriptaddress,showpacs,amssymb,prl,floatfix,longbibliography]{revtex4-2}

\usepackage{graphicx}
\usepackage[]{hyperref}
\hypersetup{colorlinks=true,linkcolor=blue,citecolor=blue,urlcolor=blue,pdfpagemode=UseNone}
\usepackage{amsmath}
\usepackage{amssymb}
\usepackage{bm}
\usepackage{xcolor}
\usepackage[normalem]{ulem}

\newcommand{\tmvo}      {TmVO$_4$}
\newcommand{\tmvox}      {Tm$_{1-x}$Y$_x$VO$_4$}

\begin{document}

\title{Nuclear magnetic resonance studies in a model transverse field Ising system}

\author{Y-H. Nian}
\affiliation{Department of Physics and Astronomy, University of California Davis, Davis, CA }

\author{I. Vinograd}
\affiliation{Department of Physics and Astronomy, University of California Davis, Davis, CA }

\author{C. Chaffey}
\affiliation{Department of Physics and Astronomy, University of California Davis, Davis, CA }

\author{Y. Li}
\affiliation{Geballe Laboratory for Advanced Materials and Department of Applied Physics, Stanford University,  CA 94305, USA}

\author{M. P. Zic}
\affiliation{ Geballe Laboratory for Advanced Materials and Department of Physics, Stanford University,  CA 94305, USA}

\author{P. Massat}
\affiliation{Geballe Laboratory for Advanced Materials and Department of Applied Physics, Stanford University,  CA 94305, USA}

\author{R. R. P. Singh}
\affiliation{Department of Physics and Astronomy, University of California Davis, Davis, CA }

\author{I. R. Fisher}
\affiliation{Geballe Laboratory for Advanced Materials and Department of Applied Physics, Stanford University,  CA 94305, USA}

\author{N. J. Curro}
\email{njcurro@ucdavis.edu}
\affiliation{Department of Physics and Astronomy, University of California Davis, Davis, CA }

\date{\today}

\begin{abstract}
The suppression of ferroquadrupolar order in TmVO$_4$ in a magnetic field is well-described by the transverse field Ising model, enabling detailed studies of critical dynamics near the quantum phase transition.  We describe nuclear magnetic resonance measurements in pure and Y-doped single crystals.  The non-Kramers nature of the ground state doublet leads to a unique form of the hyperfine coupling that exclusively probes the transverse field susceptibility. Our results show that this quantity diverges at the critical field, in contrast to the mean-field prediction.  Furthermore, we find evidence for quantum critical fluctuations present near Tm-rich regions in Y-doped crystals at levels beyond which long-range order is suppressed, suggesting the presence of quantum Griffiths phases.   
\end{abstract}
\maketitle

\section{Introduction}

Unconventional superconductivity tends to emerge in the vicinity of a quantum critical point (QCP), where some form of long-range ordered state is continually suppressed to $T=0$ \cite{balakirev2003signature,ColemanQCreview,tuson,tusonNature2008,MatsudaB122review2014,Kuo2015,KivelsonNematicQCP2015, Schuberth2016,Paschen2020}. This observation suggests that there may be an important relationship between the superconducting pairing mechanism and the strong quantum fluctuations associated with the QCP, however there are major challenges to understanding the fundamental physics at play in these systems.  In practice various approaches can be utilized to tune the ordered state to the QCP. Hydrostatic pressure or magnetic field are thermodynamic variables that are homogeneous throughout the material and can be varied continuously.  Doping, on the other hand, offers a convenient method to apply `chemical pressure' or introduce charge carriers, but can introduce electronic heterogeneity at the nanoscale which can complicate interpretation \cite{AlloulHirschfeld,ParkDropletsNature2013}.  In such cases it can be difficult to disentangle what experimental observations to ascribe to fundamental properties of a quantum phase transition versus extrinsic effects arising from the long-range effects of the dopants.

\begin{figure}[!t]
\begin{center}
    \includegraphics[width=\linewidth]{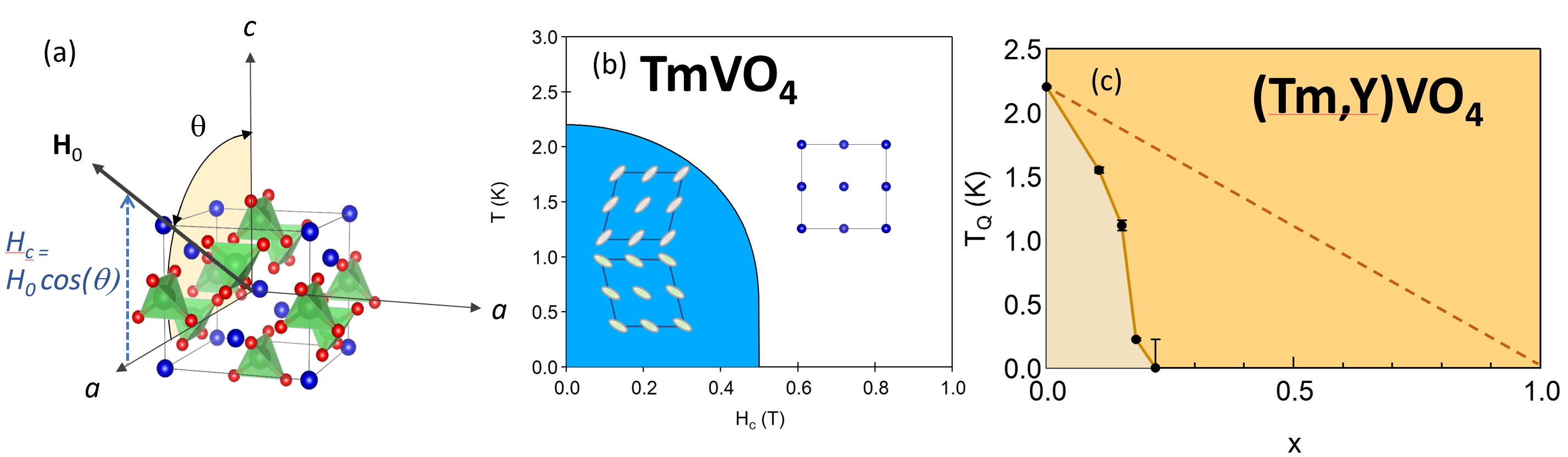}
    \caption{\label{fig:summary} (a) Crystal structure of TmVO$_4$  ($I41/amd$) with Tm atoms in blue, V atoms in green, and oxygen atoms in red.  For the studies discussed here, the magnetic field, $\mathbf{H}_0$, was rotated in the $ac$-plane, with an angle $\theta$ between $\mathbf{H}_0$ and the $c$ axis. The projection of the field along the $c$-axis is $H_0\cos\theta$. (b) Schematic phase diagram of \tmvo\ as a function of magnetic field $H_c$ along the $c$-axis, illustrating the $B_{2g}$ orthorhombic distortion in the ferroquadrupolar state. (c) Phase diagram for \tmvox, reproduced from \cite{Li2024}. The dashed line represents the mean-field result expected purely from dilution.}
\end{center}
\end{figure}

In order to better understand the influence of doping in strongly interacting system near a quantum phase transition, it is valuable to study a model system in the absence of superconductivity. TmVO$_4$ is as material that has attracted interest recently because its low temperature properties are well-described by the transverse field Ising model (TFIM), an archetype of quantum criticality \cite{Massat2021,sachdevbook}. \tmvo\ exhibits long-range ferroquadrupolar order in which the Tm $4f$ orbitals spontaneously align in the same direction, as illustrated in Fig. \ref{fig:summary}.  The Tm$^{3+}$ ions ($4f^{12}$ with $L=5$, $S=1$, $J=6$)  experience a tetragonal crystal field interaction, and the ground state is well separated by a gap of $\sim 77$ K to the lowest excited state \cite{Knoll1971,BleaneyTmVO4review}. The ground state is a non-Kramers doublet, so the first order Zeeman interaction vanishes for in-plane fields  (i.e. $g_c\sim 10$ while $g_a = g_b = 0$). This doublet can be described by a spin-1/2 pseudospin in which one component, $\sigma_z$, corresponds to a magnetic dipole moment oriented along the $c$-axis, while the other two components $\sigma_x$ and $\sigma_y$ correspond to electric quadrupole moments with $B_{2g}$ ($xy$) and  $B_{1g}$ ($x^2-y^2$) symmetry, respectively \cite{MelcherCJTEreview}. The two quadrupole moments couple bilinearly to lattice strains $\varepsilon_{xx}-\varepsilon_{yy}$ and $\varepsilon_{xy}$, which gives rise to an effective interaction between the moments and leads to a cooperative Jahn-Teller distortion at a temperature, $T_Q$ \cite{Gehring1975}. \tmvo\ spontaneously undergoes a tetragonal to orthorhombic distortion with $B_{2g}$ symmetry below $T_Q = 2.15$ K with orthorhombicity $\delta\approx 0.01$, as illustrated in Fig. \ref{fig:summary}(b). Because there are two distinct orientations of the quadrupolar moments, the ferroquadrupolar order has Ising symmetry that can be described as a coupling between neighboring pseudospins.  On the other hand, a magnetic field oriented along the $c$-axis couples to the pseudospin in a direction that is transverse to the ferroquadrupolar order \cite{FisherNematicQCP}. This field mixes the two degenerate ground state quadrupolar states, enhancing the fluctuations of the pseudospins and suppressing $T_Q$ at a quantum phase transition with critical field $H_c^*\approx0.5$ T \cite{Massat2021}. This interpretation has been strengthened by the recent observation of a quantum critical fan emerging from the QCP that extends to temperatures above $T_Q$ \cite{Nian2024}.

LiHoF$_4$ is another important material whose physics is well described by the TFIM \cite{RosenbaumQCmodel}. There are important differences, however, between LiHoF$_4$ and \tmvo.  Although the physics of both systems derives from non Kramers doublets, the former is a ferromagnet with Ho moments ordering along the $c$-axis, whereas the latter has ferroquadrupolar order with quadrupolar moments ordering in the plane.  As a result, the transverse field direction for LiHoF$_4$ is perpendicular to the $c$-axis, whereas in \tmvo\ the transverse field direction is parallel to $c$.  This fact is crucial for \tmvo\ because it also has profound consequences for the hyperfine coupling to neighboring nuclear spins and enables unique measurements of the quantum fluctuations directly.  Moreover, since the quadrupolar moments couple to strain fields, long-range order in \tmvo\ is particularly sensitive to dopants.  Therefore substituting with Y in \tmvo\ offers a unique opportunity to investigate how the quantum phase transition changes in response to the disorder and random fields introduced by the dopant atoms.  

\section{Couplings to Non-Kramers Doublet}

\subsection{Lattice interaction}


\subsubsection{Ground state Wavefunctions}

The ground state wavefunctions of the Tm in the $D_{4h}$ point group symmetry of the \tmvo\ lattice are given by:
\begin{equation}\label{eqn:gswavefunction}
  |\psi_{1,2}\rangle = \alpha_1 |\pm 5\rangle + \alpha_2|\pm 1\rangle + \alpha_3|\mp3\rangle
\end{equation}
in the $|J_z\rangle$ basis, where the $\alpha_i$ coefficients  are determined by the details of the crystal field Hamiltonian \cite{BleaneyTmVO4review,FisherNematicQCP}. Is is straightforward to show that  $J_{x,y}$ operators vanish in the subspace spanned by these states. On the other hand, there are three other operators that do not vanish:
\begin{equation}\label{eqn:pseudospin}
  J_x^2 - J_y^2 \sim \sigma_x \text{, \hspace{2cm}} J_xJ_y + J_yJ_x \sim \sigma_y \text{, \hspace{2cm} and } J_z \sim \sigma_z,
\end{equation}  
where the $\sigma_{\alpha}$ are the Pauli matrices. Physically, the first two operators represent quadrupolar moments with $B_{1g}$ and $B_{2g}$ symmetries, respectively, and the third represents a magnetic moment along the $z$ direction.  The conjugate fields to these moments are strain $\epsilon_{B1g} = \epsilon_{xx} - \epsilon_{yy}$, $\epsilon_{B2g} = \epsilon_{xy}$, and magnetic field $H_z$, respectively. Here the strain tensor is defined as $\epsilon_{ij} = (\partial u_i/\partial x_j - \partial u_j/\partial x_i)/2$, where $\mathbf{u}(\mathbf{x})$ is the displacement from the equilibrium lattice positions. 

\subsubsection{Cooperative Jahn-Teller Effect}

Because the quadrupolar moments have non-uniform charge distributions, they can interact with a strained lattice via a bilinear coupling of the form $-\eta_i\varepsilon_{i}\sigma_i$, where $\eta_i$ is an electron-lattice coupling constant. This coupling renormalizes the elastic constant, leading  to a softening in both the $B_{1g}$ and $B_{2g}$ channels, but is strongest for the $B_{2g}$ channel for \tmvo.  It can be shown that this leads to an effective coupling between the quadrupolar moments: 
\begin{equation}\label{eqn:TmCoupling}
  \mathcal{H}_{ex} = \sum_{l\neq l'} J(l - l') \sigma_y(l)\sigma_y(l')
\end{equation}
where the sum is over the lattice sites, and $J(l - l')$ is an Ising interaction between the Tm quadrupolar moments \cite{MelcherCJTEreview,Gehring1975}. The coupling depends on the details of the lattice, and because it is mediated by strain fields, it can extend well beyond just nearest neighbor sites.  This interaction leads to long-range order in the three-dimensional \tmvo\ lattice below a temperature $T_Q = 2.15$ K, with finite expectation values of $\pm\langle \sigma_y\rangle$.     This ferroquadrupolar order is accompanied by a 
$B_{2g}$ lattice distortion as illustrated in Fig. \ref{fig:summary}(b) \cite{TmVO4CJTE}. 

\subsection{Zeeman interaction}

The interaction between a non-Kramers doublet in a tetragonal environment and a magnetic field is given by:
\begin{equation}\label{eqn:zeeman}
  \mathcal{H}_Z = g_{z}\mu_B H_z \sigma_z + \frac{1}{2}(g_J\mu_B)^2 b\left[(H_x^2-H_y^2)\sigma_x + 2H_x H_y\sigma_y\right]
\end{equation}
where $H_{x,y}$ is a magnetic field along the $(x,y)$ direction, $g_J = 7/6$ for Tm$^{3+}$ and $g_{c}$ and $b$ depend on the crystal field Hamiltonian \cite{Washimiya1970}.  These parameters have been measured for \tmvo\ to be $g_{c}= 10.21$ and $b/k_B = 0.082 K^{-1}$ \cite{BleaneyTmVO4review}. Note that $\mathbf{H}$ couples quadratically in the $x$ and $y$ directions, rather than linearly for a Kramers doublet.  A field in the $z$ direction splits the doublet linearly, and acts as a \emph{transverse} field for the Ising interaction in Eq. \ref{eqn:TmCoupling}.

\subsubsection{Induced moments for perpendicular fields}

The Zeeman interaction can also be written as $\mathcal{H}_Z = \mathbf{\mu}\cdot\mathbf{H}$, where the magnetic moment along $z$ is $\mu_z= g_{||}\mu_B\sigma_z$, and the perpendicular fields $H_{x,y}$ can couple with quadrupolar moments giving rise to effective magnetic moments:
\begin{equation} 
\mu_{x,y} = \frac{1}{2}(g_J\mu_B)^2 b(H_x \sigma_{y,x} \pm H_y\sigma_{x,y}).
\end{equation}
For sufficiently low perpendicular fields, $H_{x,y} \leq 3$ T, the second order Zeeman interaction in the perpendicular direction will be less than $0.1 k_B T_Q$, and can be safely ignored.  At higher fields, 
$H_x$ and $H_y$ can also act either longitudinal or transverse fields for the Ising order, and can in fact be used to detwin the ferroquadrupolar order \cite{Vinograd2022}. 

\subsection{Transverse Field Ising Model for Ferroquadrupolar Order}

The low temperature degrees of the Tm electronic degrees of freedom are thus captured by the sum  $\mathcal{H}_{ex} + \mathcal{H}_Z$, which maps directly to the TFIM:
\begin{equation}\label{eqn:TFIM}
  \mathcal{H}_{Tm} = \sum_{l\neq l'} J(l - l') \sigma_y(l)\sigma_y(l') +   g_z\mu_B H_c\sum_{l} \sigma_z(l),
\end{equation}
where the sum is over the Tm lattice sites.  Here we have ignored the small contribution from the perpendicular component of the magnetic field.   Mean field theory predicts a QCP for a $c$-axis field of $T_Q/g_c\mu_B \approx 0.3$ T, which is close to the experimental value of  $H_c^* =0.5$ T.  Note that if there is a perpendicular field oriented such that $H_x$ or $H_y$ is zero, the system can still be described by the TFIM, because $\mathcal{H}_Z$ does not couple to the longitudinal order in pseudospin space ($\sigma_y$).  Rather, there is an effective transverse field in the $x$-$z$ plane of pseudospin space leading to a different value of the critical field \cite{Vinograd2022}.

\subsection{Coupling to Nuclear Spins}

\subsubsection{Hyperfine Coupling to $^{51}$V}

In most insulators the hyperfine coupling between a localized electron spin and a nearby nucleus arises due to the direct dipolar interaction and can be described as $\mathcal{H}_{hyp} = \mathbf{I}\cdot\mathbb{A}\cdot{\mathbf{J}}$, where $\mathbf{I}$ is the nuclear spin, $\mathbf{A}$ is the (traceless) hyperfine tensor, and $\mathbf{J}$ is the electron spin.  For temperatures well below the crystal field excitations, $\mathbf{J}$ should be replaced by the ground state pseudospin operators and $\mathbb{A}$ should be renormalized.  For a non-Kramers doublet, there can be no coupling along the $x$ or $y$ directions because the magnetic field of the nucleus does not interact with the doublet.  Rather, the hyperfine coupling has the form:
\begin{equation}
\label{eqn:hyp2}
\mathcal{H}_{hyp}=A_{zz}I_z \sigma_z+ C(H_x I_x - H_y I_y)\sigma_x + C (H_xI_y + H_yI_x)\sigma_y,
\end{equation}
where $A_{zz}$ and $C$ are constants \cite{Washimiya1970}.  In the absence of magnetic field, there is only a coupling along the $z$ direction, corresponding to the transverse field direction. To determine the values of the coupling $C$, note that  Eq. \ref{eqn:hyp2} can be re-written in terms of the effective magnetic moments:
\begin{eqnarray}\label{eqn:hyp3}
  \mathcal{H}_{hyp} &=&  \frac{2C}{(g_J\mu_B)^2 b}(I_x \mu_x + I_y \mu_y) + \frac{A_{zz}}{g_{z}\mu_B}I_z \mu_z\\
  &=&\gamma \hbar (h_x I_x + h_y I_y + h_z I_z),
\end{eqnarray}
where $h_{\alpha}$ are the hyperfine fields at the nucleus created by the Tm moments.  Using the measured values of  $h_{x}/{\mu_{x}} = -0.0336 T/\mu_B$ and $h_{z}/{\mu_{z}} = 0.0671 T/\mu_B$ obtained by comparing the Knight shift versus susceptibility, we can  identify \cite{Wang2021}:
\begin{eqnarray}
  C  &=& \frac{1}{2}\gamma\hbar (g_J \mu_B)^2 b \left(\frac{h_x}{\mu_x}\right) \approx -0.37 \mu K/T\\
  A_{zz} &=& \gamma\hbar g_z \mu_B \left(\frac{h_z}{\mu_z}\right)  \approx368 \mu K. 
\end{eqnarray}

\subsubsection{Quadrupolar coupling to $^{51}$V}
$^{51}$V has spin $I=7/2$ and a nuclear quadrupolar moment $Q = 0.052$ barns. Note that this moment is several orders of magnitude smaller than the electronic quadrupolar moment of the Tm $4f$ orbitals. Nevertheless, the extended charge distribution of the latter can contribute to the electric field gradient (EFG) tensor at the V nuclear site, which in turn couples to $Q$.   As a result, the nuclear spins can couple to the pseudospin via the nuclear quadrupolar interaction \cite{Washimiya1970}:
\begin{equation}
{\mathcal{H}_Q =  B_1(I_x^2-I_y^2)\sigma_x + B_2(I_xI_y+I_yI_x)\sigma_y  + P[3 I_z^2-I(I+1)]\mathbf{1}}.
\end{equation}
Note that $B_2=B_1$, and corresponds to a 45$^{\circ}$ rotation of the principal axes of the EFG.  The last term, $P$,  is determined by the local charge distribution in the VO$_4$ tetrahedra,  and is independent of the $4f$ orbitals. The EFG asymmetry parameter is given by $B_1\langle \sigma_x\rangle/P$, and can be measured through detailed spectral measurements as a function of angle in the ordered state. We estimate $P \approx 15$ $\mu$K and $B_1 = B_2 \approx 0.22$ $\mu$K \cite{Vinograd2022}. 

Of all the terms in $\mathcal{H}_{hyp} + \mathcal{H}_Q$, $A_{zz}$ is several orders of magnitude larger than any other, even for perpendicular fields of several tesla. Thus the coupling between the $^{51}$V and the Tm $4f$ orbitals is essentially only along the transverse field direction.   

\subsubsection{Hyperfine coupling to $^{169}$Tm}

$^{169}$Tm has a spin of $I=1/2$, and experiences a hyperfine coupling but no quadrupolar interaction.  By symmetry, the form of the hyperfine coupling must also be described by Eq. \ref{eqn:hyp3}.  In this case, however, the coupling  $A_{zz} \approx 160$ mK is nearly three orders of magnitude larger than that for the $^{51}$V due to the on-site coupling \cite{Schwab1975}.  As a result, the spin lattice relaxation rate in the paramagnetic state is so fast that the $^{169}$Tm resonance has not been observed.   
On the other hand, Bleaney and Wells reported $^{169}$Tm in the ferroquadrupolar state, where they found a large shift of the resonance frequency for fields applied in the perpendicular direction \cite{BleaneyTmVO4review}. In this case, the shift is due to the induced moments from the ordered Tm quadrupoles.  The shift exhibited a two-fold rotation symmetry as the field was rotated in the perpendicular direction, which they attributed to the second order Zeeman interaction and the induced magnetization.  The two-fold rotation reflects the orthorhombic crystal structure in the ferro{quadrupolar} state.

\section{Nuclear Magnetic Resonance Studies}

Recently several studies have been conducted of the $^{51}$V NMR in \tmvo\ in order to better understand the nature of the quantum phase transition \cite{Wang2021,Vinograd2022,Nian2024}.  In principle, one could perform zero-field NMR (or nuclear quadrupolar resonance, NQR) and gradually apply a $c$-axis field to investigate the behavior as the field is tuned to the QCP.  However, experiments below 1 MHz are difficult because the signal-to-noise ratio varies as $f^{3/2}$, where $f$ is frequency \cite{Hoult1976}.  To overcome this challenge, a perpendicular field of 3.3 T was applied along the [100] direction of the crystal (corresponding to the $x$ or $y$ directions in Eq. \ref{eqn:zeeman}), and the crystal was rotated to project a small component along the $c$-axis, as illustrated in Fig. \ref{fig:summary}(a).  

\begin{figure}[!t]
\begin{center}
    \includegraphics[width=\linewidth]{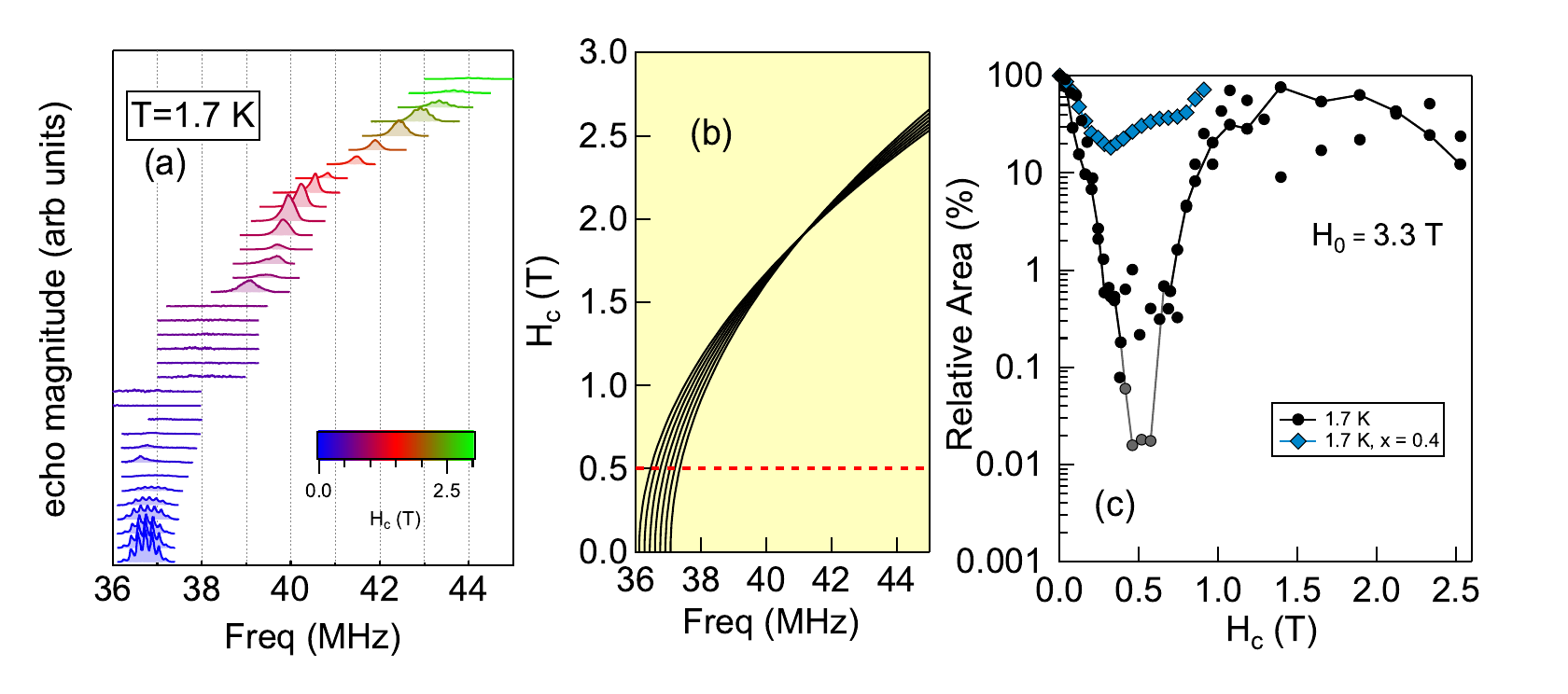}
    \caption{\label{fig:wipeout} (a) Spectra of $^{51}$V for several different values of $H_c$ as the crystal is rotated (see Fig. \ref{fig:summary}(a)). (b) Calculated frequencies of the seven transitions as a function of $H_c$.  The transitions merge at the magic angle, and then separate at higher values of $H_c$. The dashed red line corresponds to the critical field, $H_c^*$.  (c) The spectral area versus $H_c$ for several different values of temperature.  The blue diamonds correspond to \tmvox\ with $x=0.4$.}
\end{center}
\end{figure}

Spectra for several different values of $H_c$ are shown in Fig. \ref{fig:wipeout}(a).  For $H_c=0$, the spectra consist of seven transitions separated by a quadrupolar interaction $P \sim 300$ kHz, as seen in Fig. \ref{fig:wipeout}(a). As $H_c$ increases, the anisotropic Knight shift and EFG tensors alter the frequencies of the various quadrupolar satellites in a well-controlled fashion, shown in Fig. \ref{fig:wipeout}(b).  The separation between the seven peaks gradually reduces and vanishes at the magic angle (where $H_c = H_0/\sqrt{3} \approx 1.8$ T), and all the peaks shift to higher frequency, reflecting the strong magnetic anisotropy. Surprisingly, the integrated area of the spectra is dramatically suppressed in the vicinity of the QCP, as shown in Fig. \ref{fig:wipeout}(c)).  This suppression of intensity has been interpreted as evidence for quantum critical fluctuations of the transverse field, due to an increase in $T_{2}^{-1}$, the decoherence rate of the nuclear spins \cite{Nian2024}.  The size of the NMR signal, $L(t)$, depends on the time evolved, $t$, since the nuclear spins are prepared in their initial superposition state.  If $L(t)$ decays faster than the minimum time to perform an experiment, then the signal intensity will be suppressed, or `wiped out'. The data in Fig. \ref{fig:wipeout}(c) suggests that $T_2^{-1}$ reaches a maximum at the QCP.


\subsection{Transverse field susceptibility}

In general, the decay envelope, $L(t)$, of a a spin-echo can be related to the noise fluctuations of the environment.   In \tmvo, this quantity can  be written as:
\begin{equation}\label{eqn:echodecay}
  \log[L(t)/L(0)] = -\frac{ A_{zz}^2}{\hbar^2}\int_{0}^{\infty} S_{zz}(\omega) \frac{F(\omega t)}{\pi \omega^2} d\omega, 
\end{equation}
where $S_{zz}$ is the dynamical structure factor for the transverse field fluctuations:
\begin{equation}
S_{zz}(\omega) = \int_{0}^{\infty}\langle \sigma_z(\tau)\sigma_z(0)\rangle e^{i\omega\tau}d\tau,
\end{equation}
and $F(x)= 8\sin^4({x/4})$ is a filter function for the spin echo pulse sequence, which takes into account the refocusing nature of the spin echo $\pi$ pulse \cite{PulseSequenceFilterFunctionsPRB}.  The spectral area, shown in Fig. \ref{fig:wipeout}(c), is proportional to $L(t)$ at fixed $t$ corresponding to the pulse spacing in the spin echo experiment.  Because the hyperfine coupling in \tmvo\ is solely along the \emph{transverse field} direction, the nuclei are invisible to the longitudinal degrees of freedom.  Only $S_{zz}(\omega)$, the noise spectrum in the {transverse} direction,  contributes to the decoherence of the nuclear spins. This anisotropic coupling is highly unusual, but it enables us to probe the transverse fluctuations without any contamination from the longitudinal fluctuations, which diverge strongly at the QCP.  The filter function acts to remove the static or low frequency ($\omega \leq 10^5$ Hz) components of the fluctuations, which are dominated by thermal fluctuations \cite{Chen2013, Nian2024}.  The remaining contributions to $S_{zz}(\omega)$, and hence to the decay of $L(t)$, is from quantum fluctuations, which exist at finite frequency. This is because they arise from the intrinsic time evolution due to the many-body Hamiltonian, which has a finite gap except at the QCP.  The fact that $L(t)$ reaches a minimum at the QCP indicates that these quantum fluctuations are largest here.  Importantly, these extend to finite temperature, even exceeding $T_Q$.  These results thus imply that there is a broad region of phase space, a `quantum critical fan', where quantum fluctuations are present.

\begin{figure}[!h]
\begin{center}
    \includegraphics[width=\linewidth]{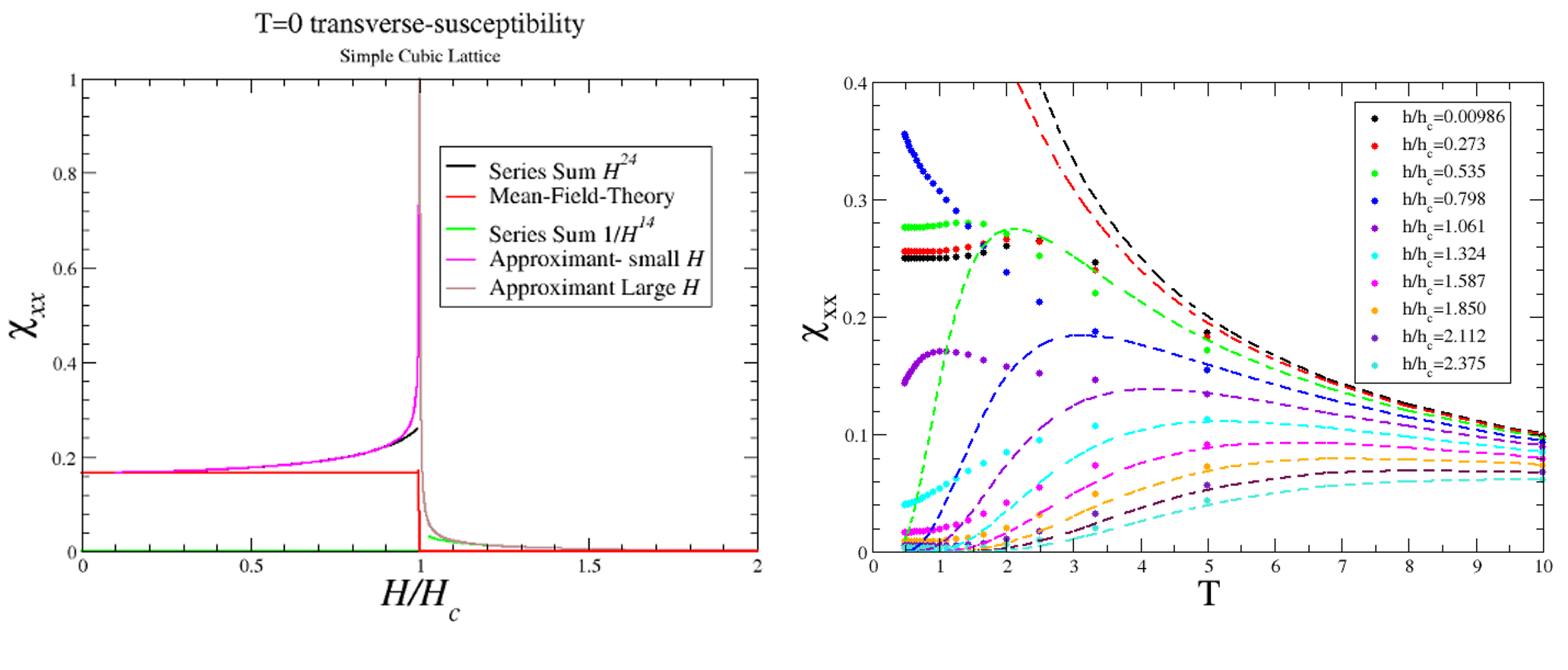}
    \caption{\label{fig:chicalc} (Left) Transverse susceptibility as a function of field for a simple cubic lattice at $T=0$ in mean-field theory and in 3D short-range models.  (Right) Temperature dependence of the transverse field susceptibility at several different values of the transverse field calculated numerically for small periodic clusters of the square-lattice.  The dashed lines are the mean field result, and the solid points of the same color are the results of numerical calculations.}
\end{center}
\end{figure}

An open question is how does the \emph{transverse} susceptibility behave in the vicinity of the quantum phase transition?  In mean-field theory at $T=0$, $\chi_{zz}$ remains constant in the ordered state, and vanishes for $H_c> H_c^*$, as shown in Fig. \ref{fig:chicalc}(a). The NMR data are inconsistent with the mean field picture, since the relative area under the spectra decreases dramatically at the QCP, indicating that  $\chi_{zz}$ must be strongly field-dependent in this range.   Numerical calculations that are based on high and low field series expansions indicate that $\chi_{zz}$ diverges logarithmically on both sides of the QCP for various 3D lattices \cite{Weihong1994}. At $T=0$ the enhancement is in a very narrow region but it should widen into a quantum critical fan at finite temperatures. Indeed we find significant differences between numerical calculations for small finite clusters and mean field theory at finite temperatures with enhancement in the general vicinity of the QCP, as seen in Fig. \ref{fig:chicalc}(b). We expect the differences to be much larger and centered at the critical point in the thermodynamic limit. These calculations, however, assume only a nearest neighbor interaction (e.g. $J(l=l') = 0$ if $l, l'$ are not nearest neighbors in Eq. \ref{eqn:TFIM}).  The interaction is expected to be long-range in \tmvo, which could tend to stabilize mean-field behavior.

\subsection{Fidelity Susceptibility}

Understanding the mechanisms of decoherence is a key problem for quantum computing, and the behavior of a central spin coupled to a well-controlled environment is an important theoretical model that has been studied extensively    \cite{Quan2006,Chen2013}.  In the case where the central spin (or qubit) is coupled to a 1D TFIM via a hyperfine coupling  along the transverse field direction, the decoherence of the qubit can be elegantly expressed in terms of the overlap of the wavefunction of the environment at different times and values of the transverse field.  In fact, the $^{51}$V spins coupled to the ferroquadrupolar ordering in \tmvo\ maps well to this  model, but with a 3D lattice for the environment  \cite{Nian2024}. Thus, \tmvo\ offers a unique opportunity to experimentally study the central spin model.

Importantly, this connection offers a new approach to understanding NMR decoherence in terms of the quantum fidelity of the environment, which is defined as the modulus of the overlap between two states: $F = |\langle \Psi'|\Psi\rangle|$.  In the case of the central spin model, the two states are $\Psi'_{\lambda}(t=0)$ and $\Psi_{\lambda + \epsilon}(t)$, where $\lambda$ corresponds to the transverse field, and $\epsilon$ corresponds to the small hyperfine field.  Two ground states of the TFIM at different values of the transverse field may initially be very similar, but will evolve strongly away from one another in the vicinity of the QCP. At $T=0$, the intensity of the NMR free induction decay is proportional to $F^2$, thus the qubit experiences a strong decoherence as the transverse field approaches the critical value.  This tendency can be captured by the fidelity susceptibility: $\chi_F = -{\partial^2F}/{\partial \epsilon ^2} $.  At finite temperatures, the fidelity can be expressed in terms of the density matrix \cite{Chen2013}. A related quantity is the Quantum Fisher Information which quantifies the sensitivity of density matrices to small changes in parameters \cite{zoller}. Because the fidelity susceptibility tends to diverge at a QCP, this quantity has been exploited theoretically to identify quantum and topological phase transitions \cite{Koenig2016,Tang2021}. 

On the surface, this picture differs from the conventional NMR picture in which decoherence arises due to  the presence of stochastic fluctuations of the hyperfine field, which can be quantitatively measured via Bloch-Wangsness-Redfield theory: $T_2^{-1} = A_{zz}^2S_{zz}(\omega=0)/2\hbar^2$ \cite{Wangsness1953,Redfield1957,CPSbook}.  However, $\chi_F$ in fact can be related to the transverse field susceptibility, $\chi_{zz} = S_{zz}/k_BT$ \cite{Gu2010}. This remarkable connection offers new insights and connections between NMR and quantum information theory.  For example, NMR wipeout is  ubiquitous in strongly correlated systems, and has been observed in the high temperature superconducting cuprates and the iron based superconductors \cite{Curro2000,Curro2000b,ImaiCuprateWipeoutPRL,JulienGlassyStripes2001PRB,Ba122ClusterGlassNMR}. In these cases, this phenomenon has been attributed to electronic inhomogeneity introduced because of the dopant atoms.  However, the behavior in \tmvo\ suggests that it might be valuable to considering the wipeout in these other systems as a consequence of their proximity to a QCP.

\section{NMR studies of Y substitution}

Replacing Tm with Y suppresses the long range ferroquadrupolar order in \tmvox\ to zero at $x_c \approx 0.22$,  as illustrated in Fig. \ref{fig:summary}(c) \cite{Li2024}. Y has no $4f$ electrons and thus lacks any magnetic or quadrupolar moments, so it acts to dilute the interactions between the Tm quadrupolar moments. The rapid suppression with doping is surprising because  mean-field theory predicts a much weaker doping dependence: $T_Q \sim 1-x$.  Y doping also suppresses ferromagnetic order in LiHoF$_4$, however in this case long-range order persists until $x = 0.95$ \cite{Babkevich2016}.  The reason for the difference between the \tmvo\ and LiHoF$_4$ is that the Y creates strain fields that couple to the ferroquadrupolar order in the former.  Y is slightly larger than Tm, thus it creates local distortions in the lattice that couple to the Tm quadrupolar moments \cite{Li2024}. This behavior is similar to that of a random field Ising model (RFIM), and causes $T_Q$ to be suppressed much faster with Y doping \cite{NATTERMANN1997}.  The local strain fields may have components with $B_{1g}$ symmetry, which couples to $\sigma_x$ and is a transverse field, as well as fields with $B_{2g}$ symmetry, which couples to $\sigma_y$ and is a longitudinal field.  

Y substitution offers an opportunity to test whether the decoherence observed in the pure \tmvo\ is due to quantum critical fluctuations.    Fig. \ref{fig:wipeout}(c) shows that for $x = 0.40$, which has no long-range ferroquadruplar order, the relative spectral area does not change significantly at $H_c^*$, in contrast to $x=0$.  This observation indicates that the quantum fluctuations are suppressed in the $x=0.40$ sample.  


\begin{figure}[!t]
\begin{center}
    \includegraphics[width=\linewidth]{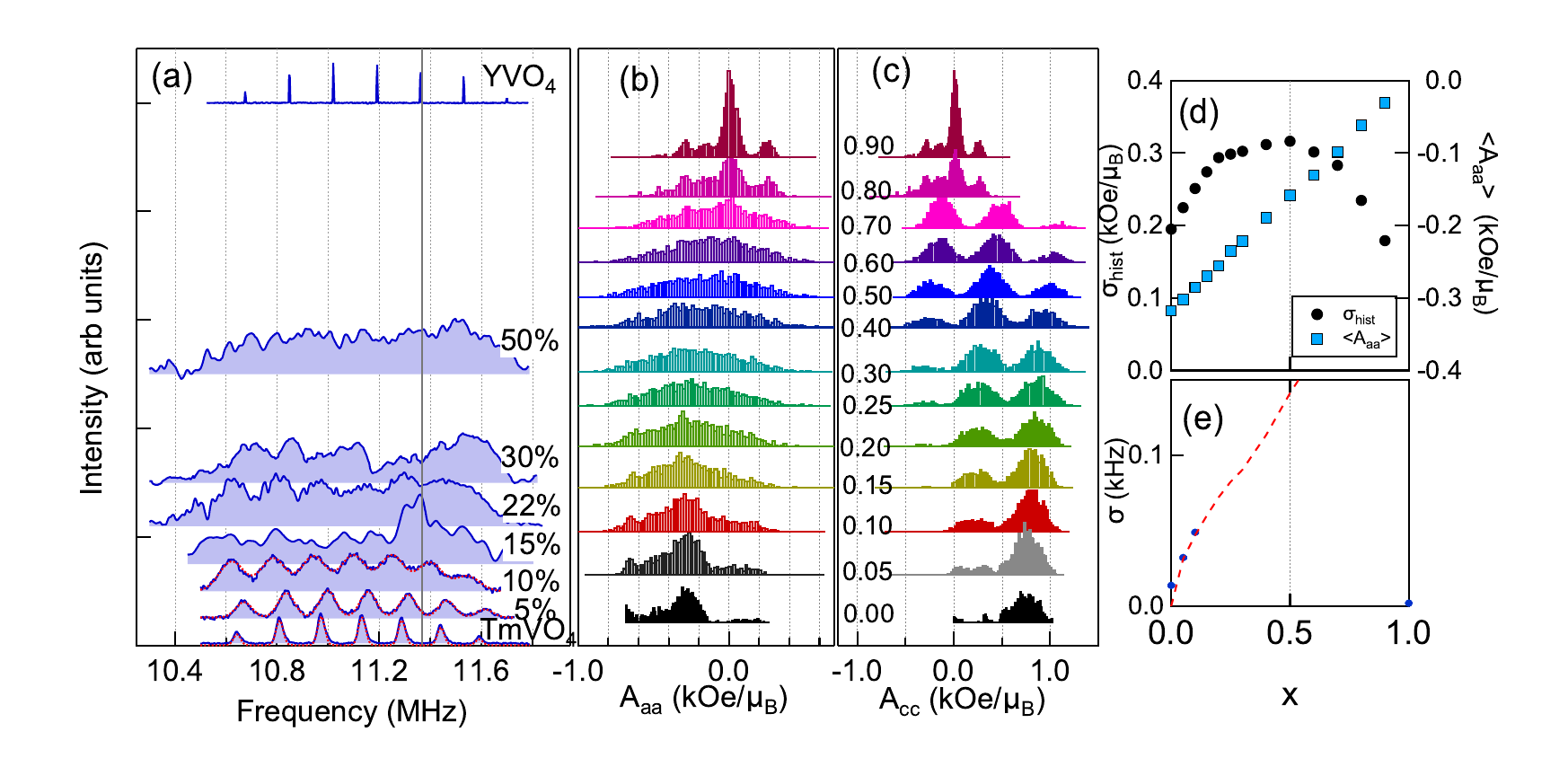}
    \caption{\label{fig:spectra} (a) Spectra for several values of $x$   measured in an external field $H_0 = 1$ T oriented perpendicular to the c-axis at 1.8 K for all but the $x=1$ case. For YVO$_4$ the spectrum was measured at 4.5 T and 10 K, but has been shifted to lower frequency by $\gamma\Delta H$ ($\Delta H = 3.9$ T) to coincide with the other spectra.  The red dotted lines are fits as described in the text. (b) and (c) Histograms of the hyperfine coupling constants, $A_{aa}$ and $A_{cc}$, respectively, for a series of Y dopings for simulations as described in the text.  (d) Average $\langle A_{aa}\rangle$ and standard deviation, $\sigma$, of the distributions shown in (b) as a function of Y doping, $x$.  (e) The measured Gaussian linewidth of the spectra shown in (a) as a function of Y doping.  The dashed red line was computed using the computed standard deviation, as discussed in the text.}
\end{center}
\end{figure}

\subsection{NMR Spectra}

NMR spectra in doped systems are generally broader than in undoped materials because the dopants generally give rise to inhomogeneity. As seen in Fig. \ref{fig:spectra}(a), the  spectra of the pure \tmvo\ and YVO$_4$ consist of seven clear resonances with small linewidths, but these resonances grow progressively broader with doping. Each of the seven resonances broadens equally between $0\leq x \leq 0.1$.  This behavior indicates that the broadening mechanism is not quadrupolar inhomogeneity, but rather a Knight shift inhomogeneity. The red dotted lines in Fig. \ref{fig:spectra}(a) are fits to the spectra, and the data in panel (e) show how the Gaussian width, $\sigma$,  varies with doping for the spectra that can be clearly fit. It is surprising that even though random strain fields are clearly present and rapidly suppressing $T_Q$,  they apparently do not significantly alter the local EFG at the V sites.  In many other strongly-correlated systems, doping usually causes significant quadrupolar broadening \cite{ImaiLSCO,GrafeNanoscaleIronPnictides,NMRnematicStrainBa122PRB2016,Menegasso2021}. In \tmvox, the larger Y atoms slightly displace the O and V in their vicinity \cite{Li2024}. On the other hand, it is possible that the VO$_4$ tetrahedra may not be significantly distorted upon Y substitution.  Also, there are two main contributions to the EFG: a lattice term arising from the arrangement of charges, and an on-site term that is determined by the electronic configuration of the local electronic orbitals \cite{CPSbook}.  It is reasonable that the latter term dominates the EFG at the V, and that the electronic configuration of the V and O orbitals remain  relatively unperturbed by Y doping.

\begin{figure}[!t]
\begin{center}
    \includegraphics[width=0.8\linewidth]{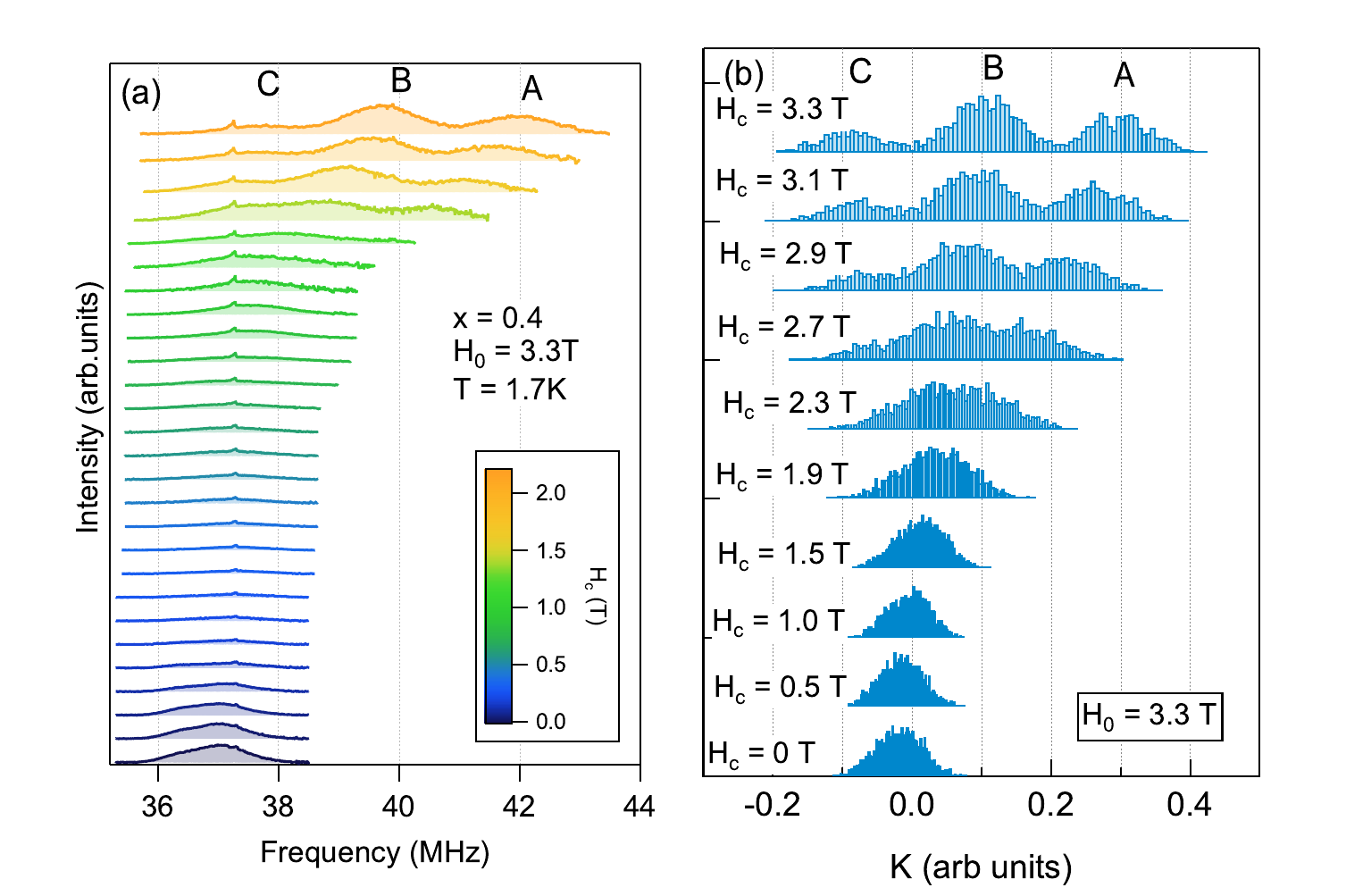}
     \caption{\label{fig:rotations} (a) Spectra of \tmvox\ with $x=0.40$ for several different values of $H_c$.  For $H_c\gtrsim 1.5$ T, three peaks are discernable, labelled $A$, $BA$, and $C$.  (b) Computed spectra based on the histograms of hyperfine couplings shown in Fig. \ref{fig:spectra}(b,c) for several different values of $H_c$ for $x=0.40$.   }
\end{center}
\end{figure}

\subsubsection{Numerical simulations}

To investigate the inhomogeneity of the magnetic environments, we computed the direct dipolar hyperfine couplings,  $A_{aa}$ and  $A_{cc}$, to the V sites in a $9\times 9\times 9$ superlattice in which a fraction of the Tm sites are randomly removed. 
Histograms of these couplings are shown in Fig. \ref{fig:spectra}(b,c) for different Y concentrations. The sum is dominated by the two nearest neighbor Tm sites along the c-axis direction (see Fig. \ref{fig:summary}(a)). The distribution  for the perpendicular direction ($A_{aa}$) broadens with doping, but does not exhibit any structure.  Fig. \ref{fig:spectra}(d) shows how the mean, $\langle A_{aa}\rangle$, and standard deviation, $\sigma_{hist}$, of the histograms vary with  Y concentration. The standard deviation increases linearly with doping, which agrees with the experimental observation of the linewidth.  The dashed red line in Fig. \ref{fig:spectra}(e) represents the expected magnetic linewidth in a field of $H_0 =1$ T, as in the experiment. This quantity is given by $\sigma(x)|K|\gamma H_0/\langle A_{aa}\rangle$, where $K = -0.66\%$.   Here we have subtracted (in quadrature) the standard deviation of the histogram of the pure TmVO$_4$ case, which includes boundary effects: $\sigma(x) =\sqrt{\sigma_{hist}^2(x) - \sigma_{hist}(0)^2}$.  The simulated linewidth agrees well with the measured linewidth, indicating that for low Y concentrations the magnetic environment of the remaining Tm is not significantly altered, despite the presence of the strain fields surrounding the Y sites.  At higher doping levels, the magnetic broadening becomes comparable to the quadrupolar splitting, and the spectra become too broad to extract any information.

\subsubsection{Effect of $c$-axis field}

Fig. \ref{fig:rotations}(a) shows how the spectra for the $x=0.40$ sample vary as the crystal is rotated in a fixed field, similar to the data shown in Fig. \ref{fig:wipeout}(a) for the $x=0$ case. As $H_c$ increases,  there is no significant wipeout at $H_c^*$, as expected since there is no long range order at this doping level and therefore no quantum critical behavior. The integrated area for these spectra are shown in Fig. \ref{fig:wipeout}(c) as a function of $H_c$. However, there are three peaks that emerge as $H_c$ increases beyond $\sim 1.5$ T, labelled $A$, $B$, and $C$, that are not present in the undoped sample.  In fact, these extra peaks are consistent with the simulated histograms of the $c$-axis hyperfine couplings shown in  Fig. \ref{fig:spectra}(c). The three peaks correspond to V sites with 0, 1 or 2 nearest neighbor Tm atoms, respectively. 

As seen in Fig. \ref{fig:spectra}(b) these different V sites should not be  discernible for a field $H_0\perp c$. On the other hand, as $\mathbf{H}_0$ rotates towards the $c$-axis, three distinct peaks should emerge.  This behavior is demonstrated in Fig. \ref{fig:rotations}(b), which displays the histograms of the Knight shift, $K(\theta) = A_{aa}\chi_{aa}\sin^2\theta + A_{cc}\chi_{cc}\cos^2\theta$, for several different values of $H_c = H_0\cos\theta$.  Here $\chi_{\alpha\alpha}$ is the static susceptibility, and we assume $\chi_{cc}/\chi_{aa} = 3$ for concreteness.   The three sites are indeed discernible for sufficiently large $H_c$, which agrees well with the observations shown in panel (a).  Moreover, the relative intensity of the peaks ($A:B:C  = 0.32:0.49:0.18$) also agrees well with the observed spectra ($0.33:0.51:0.16$).  We therefore conclude that site $A$ corresponds to V with 2 n.n. Tm, site $B$ with 1 n.n. Tm, and site $C$ with 0 n.n. Tm.  This property enables us to learn about the electronic inhomogeneity by measuring the relaxation at the different sites. 

\begin{figure}[!h]
\begin{center}
        \includegraphics[width=\linewidth]{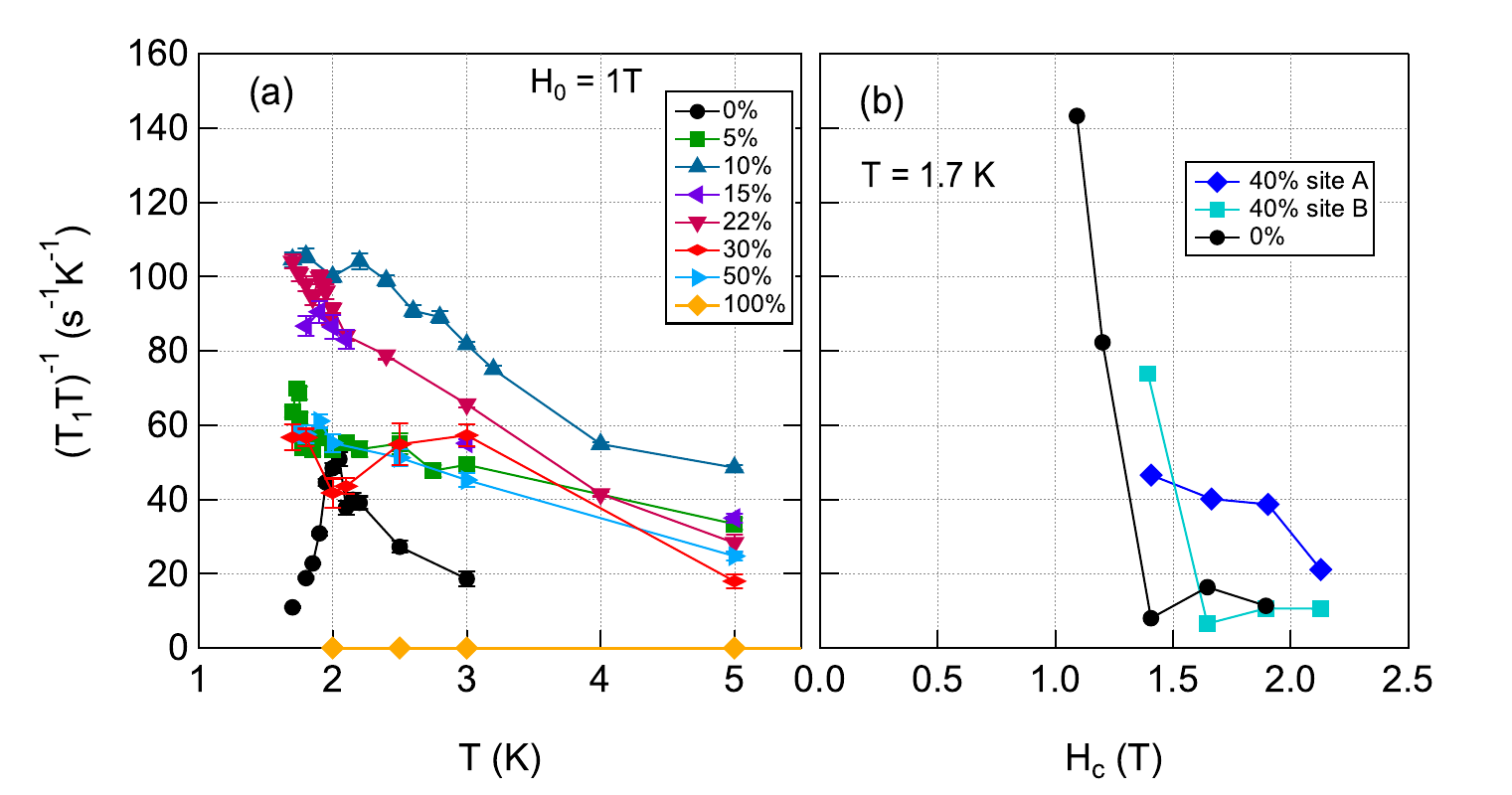}
    \caption{\label{fig:T1} (a) $(T_1T)^{-1}$ versus temperature for several different Y doping levels, measured at $H_0 = 1$ T (except for the 100\%, measured at 4.5 T), for $\theta = 90^{\circ}$.  In this case, all three sites overlap.  This data corresponds to the magnetic relaxation channel, as described in \cite{Vinograd2021}.  (b) $(T_1T)^{-1}$ versus the c-axis field component, $H_c$, for the pure \tmvo, and for the $A$ and $B$ sites in the 40\% sample.}
\end{center}
\end{figure}

\subsection{Spin Lattice Relaxation Rate}

Fig. \ref{fig:T1}(a) displays $(T_1T)^{-1}$ versus temperature for several different doping levels, measured for field perpendicular to the $c$-axis. Note that for this field orientation the resonance frequencies of sites $A$, $B$, and $C$ overlap, and thus we are unable to discern if these spin fluctuations are spatially inhomogeneous. 
There is a clear peak for the pure \tmvo\ at $T_Q$ reflecting the critical slowing down at the thermal phase transition. As the doping level increases this peak is suppressed to lower temperatures,  yet $(T_1T)^{-1}$ increases and reaches a broad maximum around $x\approx 0.10$.  In fact, the spin fluctuations appear to be enhanced near the vicinity of the critical doping level, $x_c$, possibly reflecting  quantum critical fluctuations at this doping.  At higher doping levels, the fluctuations gradually are suppressed and eventually disappear.  For the pure YVO$_4$, there are no magnetic moments present anymore, and $(T_1T)^{-1}$ is several orders of magnitude smaller.

Sites $A$, $B$ and $C$ can be discerned when there is a finite $H_c$ component present.  Fig. \ref{fig:T1}(b) compares $(T_1T)^{-1}$ versus $H_c$ in pure \tmvo\ with Tm$_{0.6}$Y$_{0.4}$VO$_4$ for the $A$ and $B$ sites.  The strong field dependence of the pure system reflects the growth of the gap as the system is tuned away from the QCP at $H_c^*$: $(T_1T)^{-1} \sim \exp(-\Delta(H_c)/T)$ \cite{Nian2024}. As $H_c$ is tuned beyond the QCP, the gap increases and $(T_1T)^{-1}$ decreases.  It is surprising that in the $x=0.40$ sample, which has no long range order, the $A$ and $B$ sites exhibit behavior that is qualitatively similar to that in the pure system.   This behavior suggests that there are still localized clusters of Tm which continue to exhibit behavior reminiscent of the undoped lattice.  Statistically there are regions of the disordered lattice with connected Tm atoms, and these may continue to exhibit correlations despite the absence of long-range order, giving rise to Griffiths phases \cite{Vojta2019}.  {An interesting open question is how such disconnected clusters may be affected by the presence of random strain fields. } 

Inhomogeneous dynamics in the disordered lattice may also explain the fact that the spectra in Fig. \ref{fig:rotations}(a) appear to exhibit an increasing intensity for $H_c\gtrsim 1.5$ T once the $A$ and $B$ peaks emerge.  If local clusters of Tm continue to exhibit quantum critical fluctuations at these sites, then $T_2^{-1}$ will be large, suppressing the signal from these sites.  In other words, the $A$ and $B$ sites may experience partial wipeout in the vicinity of $H_c^*$.  Overall these sites contribute $84\%$ of the total area, and the relative area under the spectra decreases by approximately the same value near $H_c^*$ in Fig. \ref{fig:wipeout}(c).  These observations further support the argument that the $A$ and $B$ sites are locally unperturbed by the Y dopants, and may exhibit behavior consistent with quantum Griffiths phases.

\section{Conclusions}

\tmvo\ offers a unique new experimental platform to investigate quantum critical phenomena, and to investigate the effects of doping.  The unique properties of the non-Kramers doublet in this system not only gives rise to the unusual Ising ferroquadrupolar order, but also ensures that the nuclear spins in this system only couple to the transverse field degrees of freedom.  Studies of the \tmvox\ uncovered several unexpected results.  First, despite the presence of random strain fields, the EFG at the V sites remains unperturbed, at least for low doping concentrations. As the doping level increases and the long range ferroquadrupolar order vanishes, the spin lattice relaxation rate for the V sites is enhanced, before decreasing for doping levels that exceed the critical concentration.  However, we find evidence that quantum critical fluctuations remain present for  V sites that belong to Tm-rich clusters, even beyond the critical doping level, suggesting the presence of quantum Griffiths phases in the Y-doped system. 
It is unclear whether such isolated Tm clusters also experience random transverse or longitudinal strain fields.  Further studies of this doped system will shed important light on how quantum fluctuations are destroyed by disorder.

\section*{Conflict of Interest Statement}

The authors declare that the research was conducted in the absence of any commercial or financial relationships that could be construed as a potential conflict of interest.

\section*{Author Contributions}

All crystals were synthesized by Y.L., M.Z. and I.R.F. NMR measurements were conducted by Y-H.N, I.V. and C.C.  Data analysis was carried out by I.V. and N.J.C., and numerical calculations were carried out by R.R.P.S.  The manuscript was written by N.J.C. and edited by all co-authors.

\section*{Funding}
Work at UC Davis was supported by the NSF under Grants No. DMR-1807889 and PHY-1852581, as well as the UC Laboratory Fees Research Program ID LFR-20-653926. Crystal growth performed at Stanford University was supported by the Air Force Office of Scientific Research under award number FA9550-20-1-0252. 

\section*{Acknowledgments}
We thank P. Klavins for support with cryogenic operations at UC Davis, and A. Albrecht and R. Fernandes for enlightening discussions. 





\bibliography{TmVO4papers.bib}

\end{document}